\documentclass[12pt]{article}
\usepackage{amsmath}
\usepackage{graphicx}
\usepackage{enumerate}
\usepackage{setspace}
\usepackage{natbib}
\usepackage{url} 
\usepackage{bm}
\usepackage{amssymb}
\usepackage{multirow}
\usepackage{booktabs}
\usepackage{natbib}
\usepackage[flushleft]{threeparttable}

\newcommand{\blind}{1}

\begin{document}

\def\spacingset#1{\renewcommand{\baselinestretch}%
{#1}\small\normalsize} \spacingset{1}

\def\vZ{
	\begin{pmatrix}
		1\\
		\bm{Z}_{1i}
	\end{pmatrix}}


\if1\blind
{
  \title{\bf Covariate adjustment in continuous biomarker assessment}
  \author{Ziyi Li and Yijian Huang\hspace{.2cm}\\
    Department of Biostatistics and Bioinformatics, Emory University\\
    Dattatraya Patil and Martin G. Sanda\thanks{
    	The authors gratefully acknowledge National Institutes of Health (R01CA230268 and CA113913).}\\
    Department of Urology, Emory University}
  \maketitle
} \fi

\if0\blind
{
  \bigskip
  \bigskip
  \bigskip
  \begin{center}
    {\LARGE\bf Covariate adjustment in continuous biomarker assessment}
\end{center}
  \medskip
} \fi

\begin{abstract}
	\doublespacing
Continuous biomarkers are common for disease screening and diagnosis. To reach a dichotomous clinical decision, a threshold would be imposed to distinguish subjects with disease from non-diseased individuals. Among various performance metrics for a continuous biomarker, specificity at a controlled sensitivity level (or vice versa) is often desirable for clinical utility since it directly targets where the clinical test is intended to operate. Covariates, such as age, race, and sample collection, could impact the controlled sensitivity level in subpopulations and may also confound the association between biomarker and disease status. Therefore, covariate adjustment is important in such biomarker evaluation. In this paper, we suggest to adopt a parsimonious quantile regression model for the diseased population, locally at the controlled sensitivity level, and assess specificity with covariate-specific control of the sensitivity. Variance estimates are obtained from a sample-based approach and bootstrap. Furthermore, our proposed local model extends readily to a global one for covariate adjustment for the receiver operating characteristic (ROC) curve over the sensitivity continuum. We demonstrate computational efficiency of this proposed method and restore the inherent monotonicity in the estimated covariate-adjusted ROC curve. The asymptotic properties of the proposed estimators are established. Simulation studies show favorable performance of the proposal. Finally, we illustrate our method in biomarker evaluation for aggressive prostate cancer. 
\end{abstract}

\noindent%
{\it Keywords:}  Quantile regression; Receiver operating characteristic curve; Sensitivity; Specificity; Specificity at controlled sensitivity.

\spacingset{1.5} 
\section{Introduction}
\label{sec:intro}

Continuous biomarker is often utilized for disease screening and diagnosis, where a threshold is imposed to reach a dichotomous clinical decision. For its performance assessment, the receiver operating characteristic (ROC) curve provides a comprehensive evaluation across all possible thresholds. Area under the ROC curve (AUC) is a popular performance metric, but it may not be clinically sensible \citep{hanley1982meaning}. Obviously, a continuous biomarker would not operate at all thresholds, since a diagnostic test typically needs to reach a certain sensitivity (or specificity) level to be clinically useful. Therefore, specificity at a controlled sensitivity level (or vice versa) would be a clinically more desirable performance metric. For example, for the non-invasive diagnosis of aggressive prostate cancer, the cost of a false negative is usually much higher than that of a false positive as a positive test result would be confirmed with biopsy. Thus the clinical utility of a continuous biomarker would be best measured with specificity at a controlled high sensitivity level, e.g., $95\%$ \citep{sanda2017association}.  In this work we mainly focus on specificity at a controlled sensitivity level. The same methods proposed can be directly applied to sensitivity at a controlled specificity level by switching the roles of cases and controls. 

\citet{platt2000bootstrap} and \citet{zhou2005improved}, among others, studied the estimation of such a metric, in the absence of covariates. However, within the diseased and non-diseased populations, there are usually other factors that influence a biomarker, such as age and ethnicity as well as specimen collection condition. For example, prostate-specific antigen (PSA), as a prostate cancer biomarker, tends to be higher in older men \citep{partin1996analysis}. In addition, African American men have higher PSA than men of other racial backgrounds \citep{henderson1997prostate, sanda2017association}. While intrinsically they do not discriminate diseased from non-diseased, these covariates may impact the performance of a biomarker in a number of ways \citep{pepe2003statistical}. In fact, covariates may confound the association between the biomarker and disease status when the covariate distributions differ between diseased and non-diseased individuals. Even when the two covariate distributions are the same, ignoring the covariates may lead to biased accuracy assessment. At a minimum, when a test is intended to operate at a controlled sensitivity level, covariate adjustment for the threshold would be necessary to ensure a uniform sensitivity level across subpopulations. This would also facilitate comparison of multiple biomarker studies, where the covariate distributions are likely different. 

In this paper, we develop a covariate adjustment method for specificity at a controlled sensitivity level, by adopting the quantile regression model \citep{koenker1978regression} at the given sensitivity for the diseased population. The proposal also extends readily to the continuous spectrum of sensitivity levels so as to address covariate adjustment for the ROC curve. Our model involves minimal assumptions. In the special case that the covariates have a finite number of values, the quantile regression model becomes saturated and thus does not actually impose any assumptions. Our method then reduces to the nonparametric method as considered by \citet{janes2009adjusting}. This seemingly natural model has not been favorably considered previously due to a few technique difficulties particularly in the circumstance of covariate adjustment for the ROC curve \citep{pepe2003statistical}. First, the computation burden may be of concern as the covariate effects are allowed to vary over quantiles. Second, the standard quantile regression of \citet{koenker1978regression} does not respect the inherent monotonicity of the conditional quantile functions. Subsequently, the estimated ROC after covariate adjustment may not even be monotone. Both issues are resolved in our proposal.

There are many existing methods for covariate adjustment in the assessment of continuous biomarkers. Most of them model covariate effects on biomarker as a whole, not necessarily at a specific sensitivity/specificity level. For example, \citet{tosteson1988general} and \citet{pepe1998three} modeled the covariate effects on the diseased and non-diseased populations, and then derived covariate-specific ROC curve. Some other methods directly estimate covariate-adjusted ROC curve through generalized linear regression, e.g., \citet{pepe1997regression, pepe2000interpretation}, \citet{cai2002semiparametric}. However, the covariate effect could be different at different sensitivity levels, as recognized by some of these authors. For example, \citet{cai2002semiparametric} discussed the possibility of including interactions between false-positive rates and covariates. In addition to the nonparametric estimator mentioned earlier, \citet{janes2009adjusting} also considered a semiparametric estimator based on normal linear model or location-scale model to adjust for quantile-specific covariate effects.

This paper is organized as follows. Section 2 presents the proposed covariate adjustment method for specificity at a controlled sensitivity level. Section 3 extends the proposal to global covariate adjustment over all sensitivity levels, resulting in a covariate-adjusted ROC curve. Simulation studies are presented in Section 4, and a real data illustration given in Section 5. Final discussions are provided in Section 6. Technical proofs are relegated to the Appendix.

  
\section{Specificity at a controlled sensitivity level}
Denote the case biomarker of interest by $M_{1}$ and its associated covariate by $\mathbf{Z}_1$. The case sample consists of $n_1$ i.i.d. replicates of $(M_{1}, \mathbf{Z}_1)$, $(M_{1i}, \mathbf{Z}_{1i}), i = 1, \cdots, n_1$. Similarly, denote the control biomarker by $M_0$ and its associated covariate by $\mathbf{Z}_0$, and the control sample consists of $n_0$ i.i.d. replicates $(M_{0j}, \mathbf{Z}_{0j}), j = 1, \cdots, n_0$. These covariates may be discrete or continuous. 

Write the conditional distribution function of the cases as $F_1(t; \mathbf{z}) \equiv \Pr(M_{1} \le t | \mathbf{Z}_{1} = \mathbf{z})$. The corresponding conditional quantile function is $F_1^{-1}(\cdot;\mathbf{z})$. Controlling sensitivity level at $\rho_0$, between $0$ and $1$, yields a test threshold to be the $(1 - \rho_0)$-th quantile, $F_1^{-1}(1 - \rho_0; \mathbf{z})$. We adopt the following quantile regression model for the relationship between $\rho_0$-level sensitivity and the covariates:
\begin{equation} \label{local}
F_1^{-1}(1 - \rho_0; \mathbf{z}) = (1,\mathbf{z}^T)\bm{\beta},
\end{equation}
where $\bm{\beta}$ is the regression coefficient; note 1 is added to the covariate vector to incorporate an intercept. This model imposes a structure only at the controlled sensitivity level $\rho_0$. In the case of the $K$-sample problem, this model is saturated and no model structure is actually imposed. For controls, we similarly define $F_0(t;\mathbf{z})$ as the conditional distribution function, i.e., $F_0(t; \mathbf{z}) \equiv \Pr(M_{0} \le t | \mathbf{Z}_{0} = \mathbf{z})$. Write $\bm{\beta}_0$ as the true value of $\bm{\beta}$. The pooled specificity at controlled sensitivity $\rho_0$ is given by 
$$
\phi_0 = \Pr\{M_{0} \le (1, \bm{Z}_{0}^{T})\bm{\beta}_0\} = E\big[ F_0(t;\mathbf{Z}_{0}) \big].
$$
This measure gives the overall specificity with covariate-specific threshold so as to keep the same controlled sensitivity level for covariate-specific subpopulations. 

Standard quantile regression method by \cite{koenker1978regression} gives a point estimator $\widehat{\bm{\beta}}$, which is a solution to the following estimating equation:
$$
n_1^{-1}\sum\limits_{i = 1}^{n_1} \vZ\big[I \{M_{1i} > (1, \bm{Z}_{1i}^T)\bm{\beta}\} - \rho_0\big] = O(n_1^{-1}).
$$
Then an estimator of $\phi_0$ can be obtained via the plug-in principle 
$\widehat{\phi} = n_0^{-1} \sum\limits_{j = 1}^{n_0} I\{M_{0j} \le (1,\bm{Z}_{0j}^T) \widehat{\bm{\beta}}\}$. In the special case of the $K$-sample problem, our estimator reduces to the nonparametric estimator of \citet{janes2009adjusting}. 

\subsection{Asymptotic study}
Now we consider the asymptotic properties of our proposed estimator. The following regularity conditions are imposed:

\begin{itemize}
	\item[] \textit{Condition 1.} The control and case size ratio $n_0/n_1$ approaches a constant $c > 0$ as $n_0 + n_1 \rightarrow \infty$.
	\item[] \textit{Condition 2.} Covariates $\mathbf{Z}_1$ and $\mathbf{Z}_0$ are bounded.
	\item[] \textit{Condition 3.} $E(\mathbf{\widetilde{Z}}_1^{\otimes2})$ is nonsingular, where $\mathbf{\widetilde{Z}}_1 = \big(1, \mathbf{Z}_1^T\big)^T$ and $\bm{v}^{\otimes2} = \bm{vv}^T$ for vector $\bm{v}$.
	\item[] \textit{Condition 4a.} Both $F_1(t;\mathbf{z})$ and $F_0(t;\mathbf{z})$ are differentiable at the threshold $t = (1, \bm{z}^T)\bm{\beta_0}$ with derivative bounded away from $0$ and $\infty$ uniformly in $\mathbf{z}$ over the supports of $\bm{Z}_1$ and $\bm{Z}_0$, respectively. 
\end{itemize}

\noindent These conditions are standard and mild. In particular, the differentiability assumption in Condition 4a is only imposed at the threshold of interest, whereas $F_1$ and $F_0$ could be discontinuous elsewhere.\\

\noindent \textit{Theorem 1.}  Suppose that the quantile regression model for the cases as given in (\ref{local}) holds locally at the $(1-\rho_0)$-th quantile, along with Conditions 1, 2, 3, and 4a.
Then, $\widehat{\phi}$ is consistent for $ \phi_0$. In addition, $n_0^{1/2}(\widehat{\phi} - \phi_0)$ converges to a normal distribution with mean zero and variance

\begin{equation} \label{func2}
V = c\rho_0(1-\rho_0)\bm{D}_2^T \bm{D}_1^{-1} \bm{D}_0 \bm{D}_1^{-1} \bm{D}_2 + \phi_0(1 - \phi_0),
\end{equation}
where $\mathbf{\widetilde{Z}}_0 = \big(1, \mathbf{Z}_0^T\big)^T$, $\bm{D}_0 = E\bm{\widetilde{Z}}_1^{\otimes 2}$, $\bm{D}_1 = E\{F^{'}_1(\bm{\widetilde{Z}}_1^T\bm{\beta}_0;\bm{Z}_1)\bm{\widetilde{Z}}_1^{\otimes 2}\}$, $\bm{D}_2 = E\{F^{'}_0(\bm{\widetilde{Z}}_0^T \bm{\beta}_0;\bm{Z}_0) \bm{\widetilde{Z}}_0\}$, and $F^{'}_1(\cdot;\bm{z}), F^{'}_0(\cdot;\bm{z})$ are the derivatives of $F_1(\cdot;\bm{z}), F_0(\cdot;\bm{z})$, respectively.

\subsection{Inference}

Theorem 1 provides the asymptotic variance for the proposed estimator. Since derivatives of the distribution functions are involved, direct estimation, however, is difficult. To overcome this difficulty, we adopt the method of \citet{huang2002calibration} for variance estimation with non-smooth estimating functions. Recast the estimator $(\widehat{\bm{\beta}}, \widehat{\phi})$ as the solution to the following set of estimating equations:
\begin{eqnarray}
\bm{G}_n(\bm{\nu}) = 
\begin{pmatrix}
n_1^{-1}\sum\limits_{i = 1}^{n_1} \vZ\big[I \{M_{1i} > (1, \bm{Z}_{1i}^T)\bm{\beta}\} - \rho_0\big]\\
n_0^{-1}\sum\limits_{j = 1}^{n_0} \big[I\{M_{0j} \le (1, \bm{Z}_{0j}^T)\bm{\beta}\} - \phi\big]
\end{pmatrix}
\end{eqnarray}
where $\bm{\nu} = (\bm{\beta}^T, \phi)^T$. Denote the true value by $\bm{\nu}_0 = (\bm{\beta}_0^T, \phi_0)^T$. The asymptotic variance of $(\widehat{\bm{\beta}}^T, \widehat{\phi})^T$ is $\bm{\Gamma}^{-1}\bm{\Sigma}(\bm{\Gamma}^{-1})^{T}$, where $\bm{\Sigma}$ is the asymptotic variance of $\bm{G}_n(\bm{\nu}_0)$ and $\bm{\Gamma}$ is the derivative of the limit of $\bm{G}_n(\bm{\nu})$ at $\bm{\nu}_0$. Note that $V$ in (\ref{func2}) corresponds to the last diagonal element of $n_1 \bm{\Gamma}^{-1}\bm{\Sigma}(\bm{\Gamma}^{-1})^{T}$. Sandwich variance estimation cannot be directly applied, since $\bm{G}_n(\bm{\nu})$ is not differentiable in $\bm{\beta}$. The method of \citet{huang2002calibration} resolves this issue. Specifically, start with an estimator for $\bm{\Sigma}$ as
\begin{align}
\widehat{\bm{\Sigma}} = 
\begin{pmatrix}
n_1^{-2}\sum\limits_{i = 1}^{n_1} \vZ^{\otimes 2}\big[I \{M_{1i} > (1, \bm{Z}_{1i}^T)\widehat{\bm{\beta}} \} - \rho_0\big]^2 & \bm{0} \\
\bm{0} & n_0^{-2}\sum\limits_{j = 1}^{n_0} \big[I\{M_{0j} \le (1, \bm{Z}_{0j}^T)\widehat{\bm{\beta}}\} - \widehat{\phi}\big]^2
\end{pmatrix}.
\end{align} 
Perform the Cholesky decomposition to give $\bm{\widehat{\Sigma}} = \bm{C}^{\otimes 2}$. Write $\bm{C} = (\bm{c}_1, \cdots, \bm{c}_L)$ with column vectors $\bm{c}_1, \cdots, \bm{c}_L$, where $L$ is the length of $\bm{\nu}$. Then, a sample-based variance estimator for $(\bm{\beta}_0^T, \phi_0)^T$ is given by $(\bm{G}_n^{-1}(\bm{c}_1) - \widehat{\bm{\nu}}, \cdots, \bm{G}_n^{-1}(\bm{c}_L) - \widehat{\bm{\nu}})^{\otimes 2}$. This method overcomes the non-differentiability issue discussed before effectively by a numerical differentiation of the inverse estimating equation $\bm{G}_n^{-1}(\cdot)$ using a data-adaptive bandwidth. 

For the variance estimation, a computationally more intensive alternative is bootstrap with resampling for cases and controls drawn separately. This approach has been commonly adopted for related problems (e.g., \citealt{janes2009adjusting}).

\section{ROC curve}
The preceding methods target a particular sensitivity level of interest. The same modeling strategy readily extends to each and every sensitivity level in a continuum, resulting in a covariate-adjusted ROC curve. As such, we impose the global model,
\begin{equation} \label{global}
F_{1}^{-1}(1 - \rho; \bm{z}) = (1, \bm{z}^T) \bm{\beta}_0(\rho) \qquad \forall \rho \in (0,1),
\end{equation}
where the regression coefficient function $\bm{\beta}_0(\rho)$ may vary with $\rho$. The pooled specificity also varies with $\rho$, 
\begin{equation} \label{globalrho}
\phi_0(\rho) = \Pr\{ M_{0} \le (1, \bm{Z}_{0}^{T}) \bm{\beta}_0(\rho)\}.
\end{equation} 
Clearly, the global model is a sub-model of the local one given by (\ref{local}). Nevertheless, the global model is fairly general itself as being nonparametric. Just the same as the local model, this global model actually imposes no structure whatsoever in the special case of the $K$-sample problem \citep{huang2010quantile}. 

Since the global model implies local models for each $\rho$ value in $(0, 1)$, we apply the estimation procedure described in Section 2 in a pointwise fashion to obtain the estimators. The regression coefficient estimator $\widehat{\bm{\beta}}(\rho)$ is the solution of
\begin{equation*} \label{estrho}
n_1^{-1}\sum\limits_{i = 1}^{n_1} \vZ\big[I \{M_{1i} > (1, \bm{Z}_{1i}^T)\widehat{\bm{\beta}}(\rho)\} - \rho\big] = O(n_1^{-1}).\\
\end{equation*} 
and an estimator of $\phi(\rho)$ is $\widehat{\phi}(\rho) = n_0^{-1}\sum\limits_{j = 1}^{n_0} I\{ M_{0j} \le (1, \bm{Z}_{0j}^{T}) \widehat{\bm{\beta}}(\rho)\}$. The computation might be perceived as intensive to have a solution at each and every $\rho$. Nevertheless, the estimator $\widehat{\bm{\beta}}(\rho)$ is a step function and its computation can be formulated as a parametric programming problem with $\sim n\log(n)$ breakpoints to examine (\citealt{koenkerquantile}, section 6.3). Starting from $\rho=0$ upward, this algorithm involves alternately solving the equation at the current $\rho$ value and finding the next breakpoint. The computation burden does not impose a real concern for most applications.

\subsection{Asymptotic study}

For the asymptotic study with the global model, we strengthen Condition 4a.

\begin{itemize}
	
	\item[] \textit{Condition 4b.} Both $F_1(t;\bm{z})$ and $F_0(t;\bm{z})$ have density functions $f_1(t;\bm{z})$ and $f_0(t;\bm{z})$, respectively, which are continuous in $t$ for given $\bm{z}$ and bounded uniformly in $t$ and $\bm{z}$ over the supports of $\bm{Z}_1$ and $\bm{Z}_0$, respectively. Meanwhile, $\bm{\beta}_0(\cdot)$ is continuously differentiable on $[\rho_1, \rho_2]$ for any $\rho_1$ and $\rho_2$ such that $0<\rho_1<\rho_2<1$. 
	
\end{itemize}

\noindent The existence of density function is a standard condition when ROC curve is of interest \citep{janes2009adjusting}. The differentiability of $\bm{\beta}_0(\cdot)$ is also mild and commonly imposed \citep{koenkerquantile}. Under the condition, there is no zero-density intervals and thus no jump in quantile.\\

\noindent \textit{Theorem 2.} Suppose that the quantile regression model for the cases as given in (\ref{global}) holds globally over $(1-\rho_2)$-th through $(1-\rho_1)$-th quantile for $0<\rho_1<\rho_2<1$, along with Conditions 1, 2, 3, and 4b.
Then, $\widehat{\phi}(\rho)$ converges in probability to $ \phi_0(\rho)$ uniformly over $\rho \in [\rho_1, \rho_2]$. Furthermore,  $n_0^{1/2}\{\widehat{\phi}(\rho) - \phi_0(\rho)\}$ converges weakly to a Gaussian process over $\rho\in[\rho_1,\rho_2]$.

\subsection{Monotonization of the estimated ROC curve}\label{Monosection}

As mentioned in the introduction section, lack of monotonicity in the estimated conditional quantile functions could result in that of the estimated covariate-adjusted ROC curve. In fact, a few existing works adopted location-scale models to avoid illogical results in estimating quantiles, e.g., \citet{he1997quantile} and \citet{heagerty1999semiparametric}. However, their models become more restrictive. We rather restore monotonicity in the ROC estimation under the original quantile regression model and suggest two approaches below.

The root of the issue is the lack of monotonicity-respecting with the estimated quantile regression coefficient process. \citet{huang2017restoration} developed a method to restore the monotonicity-respecting property by identifying and interpolating monotonicity-respecting breakpoints of the original estimated coefficient process. We apply this approach to obtain a monotonicity-respecting estimator $\widetilde{\bm{\beta}}(\cdot)$. Plugging this estimator $\widetilde{\bm{\beta}}(\cdot)$ in (\ref{globalrho}) results in an estimated ROC curve that is monotone. As a note, the resulting monotonized ROC curve is still a step function. The second strategy is to directly apply the method of \citet{huang2017restoration} to the estimated ROC curve $\{\widehat{\phi}(\rho),~ 0 \le \rho \le 1\}$. This method results in a piecewise-linear monotonized ROC curve. We refer these two methods as regression- and ROC-based monotonizations thereafter.

The monotonized estimators are asymptotically equivalent to the original estimators as shown in \citet{huang2017restoration}. For finite sample, the monotonized estimators may have efficiency gain. 

\subsection{Inference}

For a point on the estimated ROC curve, one may adopt the same inference procedure with the local model as described in (\ref{local}). However, note the availability of several point estimates, depending on whether a monotonized ROC is employed. Nevertheless, any choice of these estimates does not make a difference since they are all asymptotically equivalent. 

If the whole ROC curve is of interest, it is possible to construct confidence band using bootstrap. To estimate the distribution of $n_0^{1/2}\{\widehat{\phi}(\cdot) - \phi_0(\cdot)\}$, we can use the same bootstrap approach in local model except that the estimand now is functional. Denote the bootstrap estimator by $\phi^{*}(\cdot)$. The distribution of $n_0^{1/2}\{\phi^{*}(\cdot) - \widehat{\phi}(\cdot)\}$ conditioning on the data is asymptotically the same as $n_0^{1/2}\{\widehat{\phi}(\cdot) - \phi_0(\cdot)\}$. For $\rho \in [\rho_1, \rho_2]$ with $\rho_1$ and $\rho_2$ satisfying $0<\rho_1<\rho_2<1$, the $95\%$ equal-precision confidence band of $\widehat{\phi}(\rho)$ is given by 
$$
\widehat{\phi}(\rho) \pm \eta_{0.95} \text{SE}\{\widehat{\phi}(\rho) \},
$$
where $\text{SE}\{\widehat{\phi}(\rho)\}$ is the standard error of $\widehat{\phi}(\rho)$ and $\eta_{0.95}$ is the estimated $95\%$ percentile of $\sup_{\rho \in [\rho_1, \rho_2]} \big[|\phi^{*}(\rho) - \widehat{\phi}(\rho)|/\text{SE}\{\widehat{\phi}(\rho)\} \big]$. $\text{SE}\{\widehat{\phi}(\rho) \}$ is also based on bootstrap resamples.  One may construct a confidence band based on a monotonized ROC curve as in Section 3.2 in the same fashion, simply with $\widehat{\phi}(\cdot)$ replaced by the monotonized version.
%
%


\section{Simulations}
We evaluate the finite sample properties of our proposal under practical sample sizes. Assume the biomarker relies on two covariates $Z_1$ and $Z_2$, both following uniform distribution between $0$ and $1$. For cases, the biomarkers $M_1$'s are generated from formulation (\ref{local}) and $\bm{\beta}_0(\cdot)$ consists of an intercept and two slopes 
$
\bm{\beta}_0(\rho)=\big[\log\{-\log(\rho)\},~ 1-\rho, ~ (1-\rho)^2 \big].
$ 
For controls, the biomarkers $M_0$'s are generated from $N(-1-0.5 Z_1 - 0.5 Z_2, 2^2)$. The true specificities at controlled sensitivity levels $0.95$, $0.90$, $0.85$, and $0.80$ are $0.24$, $0.36$, $0.45$, and $0.52$, respectively.

Table \ref{tab:sim3} reports the performance of the proposed method in this setting, including bias, sample- and bootstrap-based standard errors, as well as the coverage probability of confidence intervals. We also present the logit transformation-based confidence interval, which is obtained by back-transforming the Wald type confidence interval of the logit-transformed $\phi_0$.  The estimation bias is very small and decreases with the increase of sample size. Both sample- and bootstrap-based standard errors are close to standard deviations. In addition, the coverage rate of confidence intervals are close to the nominal level under all scenarios. These demonstrate the favorable performance of the proposed method. We observe sample-based variance estimation has comparable performance with the bootstrap-based estimation, while the sample-based inference has advantages in computational efficiency.

As discussed in Section 3.2, the estimator $\widehat{\phi}(\cdot)$ may not respect the monotonicity of $\phi_0(\cdot)$, leading to illogical results. We implement the two monotonization methods described in Section \ref{Monosection} and evaluate their performance. Table \ref{tab:sim4} reports the bias and coverage rate related with these two methods where $\widehat{\phi}_{reg}$ and $\widehat{\phi}_{ROC}$ correspond to the estimators after adopting regression- and ROC-based monotonization methods, respectively. The confidence intervals of the monotonized estimators are constructed with the sample-based standard error estimates. We find both regression- and ROC-based methods show small bias and good coverage rate under different sample sizes and sensitivity levels. ROC-based approach generally results in better coverage probability than regression-based method. The coverage probability of ROC-based method is comparable or even better than that without applying monotonicity-restoration method (results presented in Table \ref{tab:sim3}). Lastly, applying monotonicity-restoration method may lead to smaller variance than original estimator, as shown for the ROC-based method and largely so for the regression-based method. This observation is consistent with the finding in \citet{huang2017restoration}.

We also consider the case with discrete covariates only. In this case, our proposed estimator coincides the nonparametric estimator in \citet{janes2009adjusting} as indicated in the introduction. The focus is on comparing the performance of our inference methods to \citet{janes2009adjusting}, under their simulation setup. The details of this simulation study are presented in Supplementary S1 and the results in Tables S1 and S2. Our sample-based variance estimation shows better coverage rate than the kernel density-based variance estimation in \citet{janes2009adjusting}, especially when controlled specificity is large ($\phi_0 = 0.95$ and $0.90$). The bootstrap-based variance estimation tends to perform better than the sample-based method when sample size is small. Meanwhile, the sample-based variance estimation performs reasonably well when sample size is moderate or large. 

\section{Illustration with a Clinical Study}

Data from a clinical study for aggressive prostate cancer \citep{sanda2017association} are used for illustration. This was a prospective, multi-center cohort of male participants for first-time prostate biopsy without pre-existing prostate cancer. After excluding $14$ subjects with missing values, the data consists of 150 subjects with aggressive (Gleason score $\ge 7$) prostate cancer, per biopsy, and 352 controls. The biomarker under consideration herein is prostate-specific antigen (PSA). Figure 1(a) shows the density of PSA from cases and controls. 

As mentioned in the introduction, elder men tend to have higher PSA values than younger men \citep{oesterling1993influence, lilja2008prostate}. Among the cases of our study, we observe significant ($p = 0.038$) elevation in PSA with the increase of age (Figure 1b). In addition to age, African-American men were also reported to have higher PSA than white men \citep{henderson1997prostate}. Figure 1c shows a small increase in PSA among African American cases compared to non-African American cases, although the increase is not statistically significant. In the following analysis, age and being African-American (AA) are included as covariates. 

The ROC curves with and without adjusting for covariates are shown in Figure 2a (black and blue curves). The adjusted ROC curves after regression- and ROC-based monotonization are also presented (red and purple curves). Adjusting for covariates leads to different ROC curve compared to the one without covariate adjustment. Covariate-adjusted specificity is higher than no-adjustment when sensitivity is between $0.7$ and $0.9$, and lower when that is between $0.2$ and $0.7$. Imposing the monotonicity does not make much difference. The exact specificity estimations for all the methods at controlled sensitivity levels $95\%$, $90\%$, $85\%$, and $80\%$ are reported in Table \ref{tab:real1}. Consistent with the observations in Figure 2a, the covariate-adjusted specificity is lower than no-adjusted specificity for $\rho_0 = 95\%$ but higher for $\rho_0 = 90\%, 85\% $, and $80\%$.

In contrast to pointwise confidence intervals, we also construct $95\%$ confidence bands for the covariate-adjusted ROC curves. As the confidence bands are similar for the ROC curves with or without monotonicity restorations, we only present the confidence band for the ROC curve after applying ROC-based monotonization in Figure 2b. The confidence band works well in providing inference for the whole ROC curve. 

Lastly, we present the estimated covariate-adjusted thresholds for PSA at controlled sensitivity level $95\%$ in Figure 2c. Such thresholds could be useful for physicians to identify subjects with aggressive prostate cancer. They ensure that the sensitivity is equally controlled among covariate-specific subpopulations. The covariate-adjusted threshold increases with age and is higher in African Americans, which aligns with existing understanding of these covariates.

Our implementation has excellent computational performance. With this prostate cancer data, which contain 150 diseased and 352 non-diseased samples, computing a covariate-adjusted ROC curve takes less than 1 second on a laptop computer with 4GB RAM and Intel Core i5 CPU. Computing the confidence band takes less than 3 seconds.

\section{Discussion}

Our contributions are two-fold. First, we provide a covariate adjustment approach for a clinical utility-sensible performance metric, specificity at controlled sensitivity or vice versa, with minimal modeling assumptions. Second, this method extends to covariate adjustment for the whole ROC curve, where the issues of computation and monotonicity have been addressed. The same statistical method also applies to sensitivity with controlled specificity as well, and our software package offers such an option. 

It is worthwhile to point out that our covariate-adjusted ROC curve represents the pooled specificity with covariate-adjusted threshold at a controlled sensitivity level. It is different from covariate-specific ROC curves, as considered by \citet{toledano1995regression, pepe1997regression, pepe1998three, pepe2000interpretation, cai2002semiparametric, cai2004semi}. The two serve different purposes. More recently, \citet{janes2009adjusting} developed non-parametric and semi-parametric methods to adjust for quantile-specific covariate effects. Their notion of covariate adjustment is similar to ours.

\subsection*{Proof of Theorems}

\noindent \textit{Proof of Theorem 1}.
We first establish the consistency and asymptotic normality of $\bm{\widehat{\beta}}$. These results for quantile regression have been established by, for example, Koenker (2005, section 4.1.1 and theorem 4.1) under fixed design. Although we consider random design, among other assumptions, similar arguments follow through to give the consistency of $\widehat{\bm{\beta}}$ and
\begin{equation} \label{koenker}
n_1^{1/2}(\bm{\widehat{\beta}} - \bm{\beta}_0) ~~\overset{d}\rightarrow ~~ N\Big(0, ~\rho_0(1 - \rho_0) \bm{D}_1^{-1}\bm{D}_0\bm{D}_1^{-1}\Big),
\end{equation}
where $\bm{D}_0 = E\bm{\widetilde{Z}}_1^{\otimes 2}$ and $\bm{D}_1 = E\{F^{'}_1(\bm{\widetilde{Z}}_1^T\bm{\beta}_0)\bm{\widetilde{Z}}_1^{\otimes 2}\}$.

Next we turn to $\widehat{\phi}$. By Condition 4a and the consistency of $\widehat{\bm{\beta}}$, the consistency of $\widehat{\phi}$ can be easily established. For asymptotic normality, we have
\begin{align*}
n_0^{1/2}(\widehat{\phi} - \phi_0)
&= n_0^{-1/2} \sum\limits_{i = 1}^{n_0} \{I(M_{0i} \le \bm{\widetilde{Z}}_{0i}^T \bm{\widehat{\beta}}) - \Pr(M_0 \le \bm{\widetilde{Z}}_{0}^T \bm{\beta}_0)\} \\
& = n_0^{-1/2} \sum\limits_{i = 1}^{n_0} \{ I(M_{0i} \le \bm{\widetilde{Z}}_{0i}^T \bm{\widehat{\beta}}) - \Pr(M_0\leq\widetilde{\bf Z}_0^T\bm{\beta} | \bm{\beta} = \bm{\widehat{\beta}}) \} \\
&~~~~~~~ + n_0^{1/2} \{ \Pr(M_0 \le \bm{\widetilde{Z}}_{0}^T \bm{\beta} | \bm{\beta} = \bm{\widehat{\beta}}) - \Pr(M_0 \le \bm{\widetilde{Z}}_{0}^T \bm{\beta}_0) \}\\
& \equiv A_n(\bm{\widehat{\beta}}) + B_n.
\end{align*}
Since $F_0(t; \bm{z})$ is differentiable at $\bm{\widetilde{Z}}_0^T\bm{\beta}_0$, in light of (\ref{koenker}), Delta method leads to
\begin{equation} \label{bndist}
B_n ~~\overset{d}\rightarrow~~ N\bigg(0,  ~c\rho_0(1-\rho_0)\bm{D}_2^T \bm{D}_1^{-1} \bm{D}_0 \bm{D}_1^{-1} \bm{D}_2 \bigg),
\end{equation}
where $\bm{D}_2 = E\{F^{'}_0(\bm{\widetilde{Z}}_0^T \bm{\beta}_0) \bm{\widetilde{Z}}_0\}$. Meanwhile, $A_n(\bm{\widehat{\beta}})$ can be written as 
\begin{equation} \label{an}
A_n(\bm{\beta}_0) + \{A_n(\bm{\widehat{\beta}}) - A_n(\bm{\beta}_0) \}, 
\end{equation}
where $A_n(\bm{\beta}_0) = n_0^{-1/2} \sum\limits_{i = 1}^{n_0} \{ I(M_{0i} \le \bm{\widetilde{Z}}_{0i}^T \bm{\beta}_0)- \phi_0 \}$. By central limit theorem,
\begin{equation} \label{andist}
A_n(\bm{\beta}_0) ~\overset{d}\rightarrow~ N\big(0, \phi_0(1-\phi_0)\big).
\end{equation}
On the other hand,
\begin{align*}
E[\{A_n(\bm{\widehat{\beta}}) - A_n(\bm{\beta}_0)\}^2] &= n_0^{-1} \sum\limits_{i = 1}^{n_0} E \{ I(M_{0i} \le \bm{\widetilde{Z}}_{0i}^T \bm{\widehat{\beta}})- I(M_{0i} \le \bm{\widetilde{Z}}_{0i}^T \bm{\beta}_0)  \\
& ~~~~~~~ -  \Pr(M_0 \le \bm{\widetilde{Z}}_{0}^T \bm{\widehat{\beta}} | \bm{\widehat{\beta}}) +  \Pr(M_0 \le \bm{\widetilde{Z}}_{0}^T \bm{\beta}_0) \}^2 \\
& = E\Big(E \big[ \{I(M_{0} \le \bm{\widetilde{Z}}_{0}^T \bm{\widehat{\beta}})- I(M_{0} \le \bm{\widetilde{Z}}_{0}^T \bm{\beta}_0) \\
&~~~~~~~ -  F_0( \bm{\widetilde{Z}}_{0}^T \bm{\widehat{\beta}}) -  F_0(\bm{\widetilde{Z}}_{0}^T \bm{\beta}_0) \}^2 | \bm{\widetilde{Z}}_{0}^T\bm{\widehat{\beta}}\big]\Big)\\
& \le E | I(M_{0} \le \bm{\widetilde{Z}}_{0}^T \bm{\widehat{\beta}})- I(M_{0} \le \bm{\widetilde{Z}}_{0}^T \bm{\beta}_0)| \\
& \le E |F_0(\bm{\widetilde{Z}}_{0}^T \bm{\widehat{\beta}})-F_0(\bm{\widetilde{Z}}_{0}^T \bm{\beta}_0)|.
\end{align*}
By Markov's inequality,
$A_n(\bm{\widehat{\beta}}) - A_n(\bm{\beta}_0) \overset{d}\rightarrow  0 $.\\
Together with (\ref{bndist}) and (\ref{andist}), Slutsky's theorem yields the result. $\square$

\vspace{0.15in}
\noindent \textit{Proof of Theorem 2.} Start with the cases, and write
\begin{align*}
\Psi_n(\bm{\beta}, \rho) &= n_1^{-1} \sum\limits_{i = 1}^{n} \bm{\widetilde{Z}}_{1i} \{I(M_{1i} > \bm{\widetilde{Z}}_{1i}^T \bm{\beta}) - \rho\}, \\
\Psi(\bm{\beta}, \rho) &= E\big[ \bm{\widetilde{Z}}_{1} \{I(M_{1} > \bm{\widetilde{Z}}_{1}^T \bm{\beta}) - \rho\} \big].
\end{align*}
It is known that $\{I(M_1 > \bm{\widetilde{Z}}_1^T \bm{\beta}): \bm{\beta \in \mathbb{R}^p}\}$ is Donsker (e.g. \citealt{kosorok2007introduction}, lemma 9.12). Furthermore, $\bm{\widetilde{Z}}_{1}$ is bounded by Condition 2. By permanence property of the Donsker class, $\{\bm{\widetilde{Z}}_1I(M_1 > \bm{\widetilde{Z}}_1^T\bm{\beta}): \bm{\beta \in \mathbb{R}^p}\}$ is Donsker. Since Donsker implies Glivenko-Cantelli, it follows that, almost surely
$$
\sup_{\bm{\beta}, \rho \in [\rho_1, \rho_2]} ||\Psi_n(\bm{\beta}, \rho) - \Psi(\bm{\beta}, \rho)|| = o(1).
$$
Thus, $||\Psi\{\bm{\widehat{\beta}}(\rho), \rho\}|| \le ||\Psi_n\{\bm{\widehat{\beta}}(\rho), \rho\}|| + ||\Psi_n\{\bm{\widehat{\beta}}(\rho), \rho\} - \Psi\{\bm{\widehat{\beta}}(\rho), \rho\}||$ leads to, almost surely,
$$
\sup_{\rho \in [\rho_1, \rho_2]} ||\Psi\{\bm{\widehat{\beta}}(\rho), \rho\}|| = o(1).
$$
It remains to be shown that, for any $\epsilon > 0$, there exists $\delta > 0$ such that if
$\sup\limits_{\rho \in [\rho_1, \rho_2]} ||\Psi\{\bm{\beta}(\rho), \rho\}|| < \delta,
\text{~then} \sup\limits_{\rho \in [\rho_1, \rho_2]}||\bm{\beta}(\rho) - \bm{\beta}_0(\rho)|| < \epsilon$.  Suppose that this is not true. Thus, for each $\delta>0$, there exists $(\zeta, \nu)$ such that $||\Psi(\zeta, \nu) - \Psi\{\bm{\beta}_0(\nu), \nu\}||< \delta$ and $||\zeta - \bm{\beta}_0(\nu)||>c$ for some constant $c > 0$. Then, there exists a subsequence of $(\zeta, \nu)$ that converges to, say, $(\zeta_0, \nu_0)$, which implies that $\zeta_0 \neq \bm{\beta}_0(\nu_0)$ also solves $\Psi(\bm{\beta}, \nu_0)$. This contradicts the fact that $\bm{\beta}_0(\rho)$ is the unique solution of $\Psi(\beta, \rho)$ for all $\rho \in [\rho_1, \rho_2]$ , as guaranteed by Condition 3 and 4a. Therefore,
$$
\sup_{\rho \in [\rho_1, \rho_2]} || \bm{\widehat{\beta}}(\rho) - \bm{\beta}_0(\rho)|| = o(1)
$$
almost surely. 

In light of the above Donsker result, for given $\rho$, $n_1^{1/2}\{\Psi_n(\bm{\beta}, \rho) - \Psi(\bm{\beta}, \rho)\}$ converges weakly to a Gaussian process. Under Conditions 2 and 4b, $n_1^{1/2}\{\Psi_n(\bm{\beta}, \rho) - \Psi(\bm{\beta}, \rho)\}$ is asymptotically uniformly equicontinuous in probability using arguments similar to \citet{huang2017restoration}, appendix. Thus, for any positive sequence $d_n = o(1)$, 
$$
\sup_{||\bm{\beta} - \bm{\beta^{'}}|| < d_n,~\rho \in [\rho_1, \rho_2]} n_1^{1/2} || \Psi_n(\bm{\beta}, \rho) - \Psi_n(\bm{\beta^{'}}, \rho) - \Psi(\bm{\beta}, \rho) + \Psi(\bm{\beta^{'}}, \rho) || = o_p(1);
$$
note that the above expression does not actually involve $\rho$. Therefore,
$$
\sup_{\rho \in [\rho_1, \rho_2]} ||\Psi_n\{\bm{\beta}_0(\rho), \rho\} + \Psi\{\bm{\hat{\beta}}(\rho), \rho\}|| = o_p(n_1^{-1/2}).
$$
Under Condition 4b, by component-wise Taylor expansion, one can show that, almost surely,
$$
\sup_{\rho \in [\rho_1, \rho_2]} \frac{|| \Psi\{\bm{\hat{\beta}}(\rho), \rho\} + E[\bm{\widetilde{Z}}_{1}^{\otimes 2}f_1\{\bm{\widetilde{Z}}_{1}^T \bm{\beta}_0(\rho) \bm{\widetilde{Z}}_{1}\}] \{\bm{\widehat{\beta}}(\rho) - \bm{\beta}_0(\rho) \} ||}{|| \bm{\widehat{\beta}}(\rho) - \bm{\beta}_0(\rho) ||} = o(1).
$$
Thus,
$$
n_1^{1/2}\{\bm{\widehat{\beta}}(\rho) - \bm{\beta}_0(\rho)  \} = n_1^{1/2} \big(E[\bm{\widetilde{Z}}_{1}^{\otimes 2} f_1\{\bm{\widetilde{Z}}_{1}^T \bm{\beta}_0(\rho) \bm{\widetilde{Z}}_{1}\}]\big)^{-1} \Psi_n\{\bm{\beta}_0(\rho), \rho\} + o_p(1),
$$
uniformly in $\rho \in [\rho_1, \rho_2]$. Therefore, $n_1^{1/2} \{\bm{\widehat{\beta}}(\cdot) - \bm{\beta}(\cdot)\}$ over $[\rho_1, \rho_2]$ converges weakly to a Gaussian process.

Now, we turn to the controls. Write $\Gamma_n(\bm{\beta}) = n_0^{-1} \sum\limits_{j = 1}^{n_0} I(M_{0j} \le \bm{\widetilde{Z}}_{0j}^T \bm{\beta})$ and $\Gamma(\bm{\beta}) = \Pr(M_{0} \le \bm{\widetilde{Z}}_{0}^T \bm{\beta})$. Similar arguments as above give
$$
\sup_{\bm{\beta}} |\Gamma_n(\bm{\beta}) - \Gamma(\bm{\beta})| = o(1).
$$
Thus, $$\sup_{\rho \in [\rho_1, \rho_2]} |\widehat{\phi}(\rho) - \phi_0(\rho)| \le \sup_{\rho \in [\rho_1, \rho_2]} |\Gamma\{\bm{\widehat{\beta}}(\rho)\} - \Gamma\{\bm{\beta}_0(\rho) \} | + o(1) = o(1)$$
almost surely, given the strong consistency of $\bm{\widehat{\beta}}(\cdot)$ and the continuity of $\Gamma(\bm{\beta})$. To establish the weak convergence of $\widehat{\phi}(\rho)$, one can show that, for any positive sequence $d_n = o(1)$,
$$
\sup_{||\bm{\beta} - \bm{\beta}^{'}|| < d_n} n_0^{1/2} |\Gamma_n(\bm{\beta}) - \Gamma_n(\bm{\beta}^{'}) - \Gamma(\bm{\beta}) + \Gamma(\bm{\beta}^{'})| = o_p(1),
$$
using similar arguments as for the cases. Therefore
\begin{align*}
n_0^{1/2}\{\widehat{\phi}(\rho) - \phi_0(\rho) \} & = n_0^{1/2} \big[\Gamma_n\{\bm{\widehat{\beta}}(\rho)\} - \Gamma\{\bm{\beta}_0(\rho) \} \big] \\
& = n_0^{1/2}\big[\Gamma_n\{\bm{\beta}_0(\rho)\} - \Gamma\{\bm{\beta}_0(\rho)\} \big] + n_0^{1/2} \big[\Gamma\{\bm{\widehat{\beta}}(\rho)\} - \Gamma\{\bm{\beta}_0(\rho)\}\big] + o_p(1) \\
& = n_0^{1/2} \big[\Gamma_n\{\bm{\beta}_0(\rho)\} - \Gamma\{\bm{\beta}_0(\rho)\}\big] + n_0^{1/2} \Gamma^{'}\{\bm{\beta}_0(\rho)\} \{\bm{\widehat{\beta}}(\rho) - \bm{\beta}_0(\rho) \} + o_p(1)
\end{align*}
uniformly in $\rho \in [\rho_1, \rho_2]$. Then, the weak convergence of $\widehat{\phi}(\rho)$ follows. $\square$

\bigskip
\begin{center}
{\large\bf SUPPLEMENTARY MATERIAL}
\end{center}

\begin{description}

\item[R-package for  caROC:] R-package caROC containing code to perform biomarker evaluation using local and global models described in the article. The package is freely available through GitHub page (link masked per competition committee request). (GNU zipped tar file) 

\end{description}



\bibliographystyle{chicago}

\bibliography{document}

\begin{table}
	\caption{Results of the simulation study for estimating specificity $\phi_0$ under controlled sensitivity level $\rho_0$. }
	\label{tab:sim3}
	\vspace{0.1in}
	\centering
	\begin{threeparttable}
		\begin{tabular}{r r r   r r r r   r r r r}
			\toprule
			\multirow{2}{*}{$n_1 = n_0$}& \multirow{2}{*}{Bias \tnote{}} & \multirow{2}{*}{SD} & & \multicolumn{3}{c}{Sample-based} & & \multicolumn{3}{c}{Boostrap-based} \\
			& & & & SE & Cov & LCov  & & SE & Cov & LCov \\
			
			\multicolumn{11}{c}{$\rho_0 = 0.95,~\phi_0 = 0.24$} \\
			100 & 191 & 799 & & 1102 & 93.82 & 94.84 & & 817 & 94.14 & 93.20 \\
			200 & 84 & 569 & & 744 & 94.38 & 95.14 & & 600 & 95.62 & 95.12 \\
			500 & 33 & 362 & & 439 & 94.48 & 94.80 & & 383 & 95.50 & 95.38 \\
			1000 & 17 & 263 & & 299 & 94.00 & 94.30 & & 270 & 94.94 & 95.20 \\
			
			\multicolumn{11}{c}{$\rho_0 = 0.90,~\phi_0 = 0.36$} \\
			100 & 65 & 760 & & 1028 & 95.22 & 96.30 & & 823 & 95.94 & 97.04 \\
			200 & 36 & 545 & & 690 & 95.24 & 95.80 & & 585 & 95.44 & 96.16 \\
			500 & 15 & 344 & & 408 & 95.40 & 95.80 & & 367 & 95.70 & 95.78 \\
			1000 & -2 & 252 & & 279 & 94.80 & 95.00 & & 257 & 94.88 & 95.16 \\
			
			\multicolumn{11}{c}{$\rho_0 = 0.85,~\phi_0 = 0.45$} \\
			100 & 20 & 725 & & 942 & 95.42 & 96.22 & & 788 & 95.86 & 96.68 \\
			200 & 23 & 529 & & 634 & 95.50 & 95.90 & & 555 & 95.52 & 96.10 \\
			500 & 4 & 330 & & 376 & 95.54 & 95.80 & & 346 & 95.74 & 95.86 \\
			1000 & 5 & 235 & & 258 & 95.44 & 95.54 & & 243 & 95.36 & 95.52 \\
			
			\multicolumn{11}{c}{$\rho_0 = 0.80,~\phi_0 = 0.52$} \\
			100 & -3 & 695 & & 878 & 95.58 & 96.48 & & 750 & 95.36 & 96.80 \\
			200 & 5 & 496 & & 589 & 95.84 & 96.36 & & 525 & 95.56 & 96.12 \\
			500 & 3 & 318 & & 352 & 95.10 & 95.24 & & 326 & 95.12 & 95.34 \\
			1000 & 1 & 222 & & 242 & 95.52 & 95.54 & & 228 & 95.14 & 95.26 \\
			\toprule
			
		\end{tabular}
		\begin{tablenotes}[normal,flushleft]
			\item {\footnotesize Bias, $(\widehat{\phi} - \phi_0)\times10^4$; SD, standard deviation ($\times10^4$); SE, mean standard error ($\times10^4$); Cov (\%) and LCov (\%), coverage rates of $95\%$ confidence interval and logit transformation-based confidence interval.}
		\end{tablenotes}
	\end{threeparttable}
\end{table}

\begin{table}
	\caption{Comparison of two monotonization methods in the simulation study. }
	\label{tab:sim4}
	\vspace{0.1in}
	\centering
	\begin{threeparttable}
		\begin{tabular}{r r    r r r r   r  r r r r}
			\toprule
			\multirow{2}{*}{$n_1 = n_0$} & & \multicolumn{4}{c}{$\widehat{\phi}_{reg}$} & & \multicolumn{4}{c}{$\widehat{\phi}_{ROC}$} \\
			& & Bias & SD & Cov & LCov & & Bias & SD & Cov & LCov \\
			
			\multicolumn{11}{c}{$\rho_0 = 0.95,~\phi_0 = 0.24$} \\
			100 & & 419 & 973 & 88.82 & 88.64 & & 276 & 776 & 94.68 & 94.58 \\
			200 & & 84 & 616 & 92.96 & 93.48 & & 154 & 562 & 94.76 & 94.94 \\
			500 & & 27 & 363 & 94.52 & 95.24 & & 56 & 363 & 94.58 & 94.70 \\
			1000 & & 26 & 259 & 94.94 & 95.18 & & 34 & 260 & 95.02 & 95.10 \\
			
			\multicolumn{11}{c}{$\rho_0 = 0.90,~\phi_0 = 0.36$} \\
			100 & & 52 & 810 & 94.44 & 95.44 & & 131 & 768 & 95.00 & 95.96 \\
			200 & & 24 & 556 & 94.90 & 95.60 & & 69 & 551 & 95.30 & 95.82 \\
			500 & & 20 & 349 & 95.22 & 95.34 & & 27 & 350 & 95.02 & 95.18 \\
			1000 & & 16 & 246 & 95.14 & 95.28 & & 18 & 248 & 95.08 & 95.12 \\
			
			\multicolumn{11}{c}{$\rho_0 = 0.85,~\phi_0 = 0.45$} \\
			100 & & 6 & 732 & 95.66 & 96.46 & & 66 & 724 & 95.86 & 96.72 \\
			200 & & 24 & 525 & 95.22 & 95.78 & & 43 & 526 & 95.46 & 96.00 \\
			500 & & 14 & 328 & 95.32 & 95.56 & & 16 & 329 & 95.24 & 95.54 \\
			1000 & & 10 & 231 & 95.44 & 95.52 & & 10 & 232 & 95.20 & 95.34 \\
			
			\multicolumn{11}{c}{$\rho_0 = 0.80,~\phi_0 = 0.52$} \\
			100 & & -14 & 685 & 95.92 & 96.68 & & 22 & 685 & 96.08 & 96.96 \\
			200 & & 13 & 495 & 95.96 & 96.40 & & 21 & 499 & 95.74 & 96.16 \\
			500 & & 5 & 311 & 95.82 & 95.98 & & 5 & 312 & 95.80 & 95.94 \\
			1000 & & 7 & 218 & 95.64 & 95.74 & & 7 & 218 & 95.82 & 95.90 \\
			\toprule
		\end{tabular}
		\begin{tablenotes}[normal,flushleft]
			\item {\footnotesize $\widehat{\phi}_{reg}$, the estimator with regression-based monotonization; $\widehat{\phi}_{ROC}$, the estimator with ROC-based monotonization;  Bias, $(\widehat{\phi}_{\cdot} - \phi_0)\times 10^4$; SD, standard deviation $\times 10^4$; Cov (\%) and LCov (\%), coverage rates of $95\%$ confidence interval and logit transformation-based confidence interval.}
		\end{tablenotes}
	\end{threeparttable}
\end{table}

\begin{table}[htp!]
	\renewcommand{\arraystretch}{1.2}
	\caption{Estimated specificity along with $95\%$ confidence interval at controlled sensitivity level ($\rho_0$) in the aggressive prostate cancer application.  }
	\label{tab:real1}
	\vspace{0.1in}
	\centering
	\begin{threeparttable}
		\begin{tabular}{c c  c c c c c c c}
			\bottomrule
			
			$\rho_0$ & & $\widehat{\phi}_{n}$ & & $\widehat{\phi}$ & & $\widehat{\phi}_{reg}$ & & $\widehat{\phi}_{ROC}$ \\ 
			0.95 & &	0.239 & &	0.196 (0.108, 0.330) & &	0.208	(0.118, 0.340) & & 0.218 (0.127, 0.348) \\
			0.90 & &	0.284 & &	0.318 (0.225, 0.429) &	& 0.316	(0.223, 0.427) & & 0.321 (0.228, 0.431) \\
			0.85 & &	0.332 & &	0.361 (0.297, 0.430) &	& 0.358	(0.294, 0.428) & & 0.361 (0.297, 0.430) \\
			0.80 & &	0.392 & &	0.398 (0.320, 0.481) &	& 0.398	(0.320, 0.480) & & 0.400 (0.323, 0.483) \\
			
			\toprule
		\end{tabular}
		\begin{tablenotes}
			\item{$\widehat{\phi}_{n}$, estimated specificity without covariate adjustment; $\widehat{\phi}$, estimated specificity adjusting for age and AA without applying monotonicity-restoration; $\widehat{\phi}_{reg}$ and $\widehat{\phi}_{ROC}$, covariate-adjusted specificity after applying regression- and ROC-based monotonization methods, respectively. The confidence interval is logit transformation-based with sample-based standard error.}
		\end{tablenotes}
	\end{threeparttable}
\end{table}

\begin{figure}[htp!]	
	\includegraphics[scale=0.71]{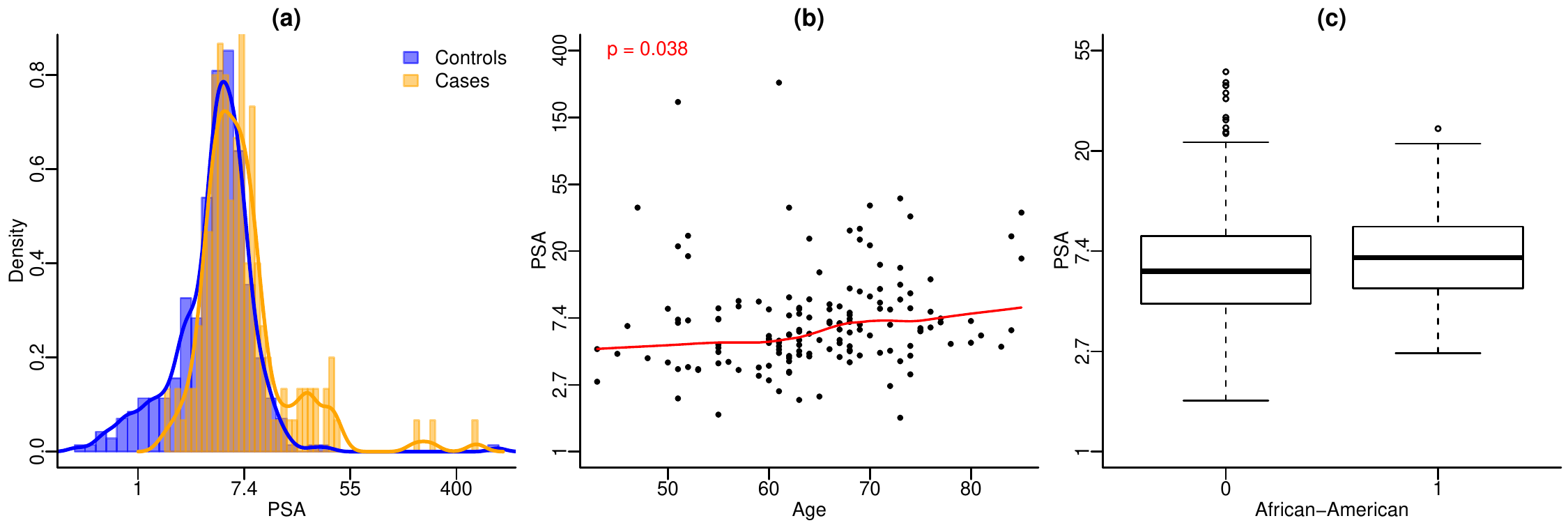}
	\caption{Exploratory plots for the clinical study. Panel (a): Histograms of PSA for cases and controls with density curves overlaid. Panel (b): Scatterplot of PSA versus age, cases only. The red solid line is fitted by loess and the p value is obtained from testing zero Pearson's correlation coefficient. Panel (c): Boxplot of PSA of African American population and non-African American group, cases only. }
	\label{overview}
\end{figure}

\begin{figure}[htp!]	
	\includegraphics[scale=0.71]{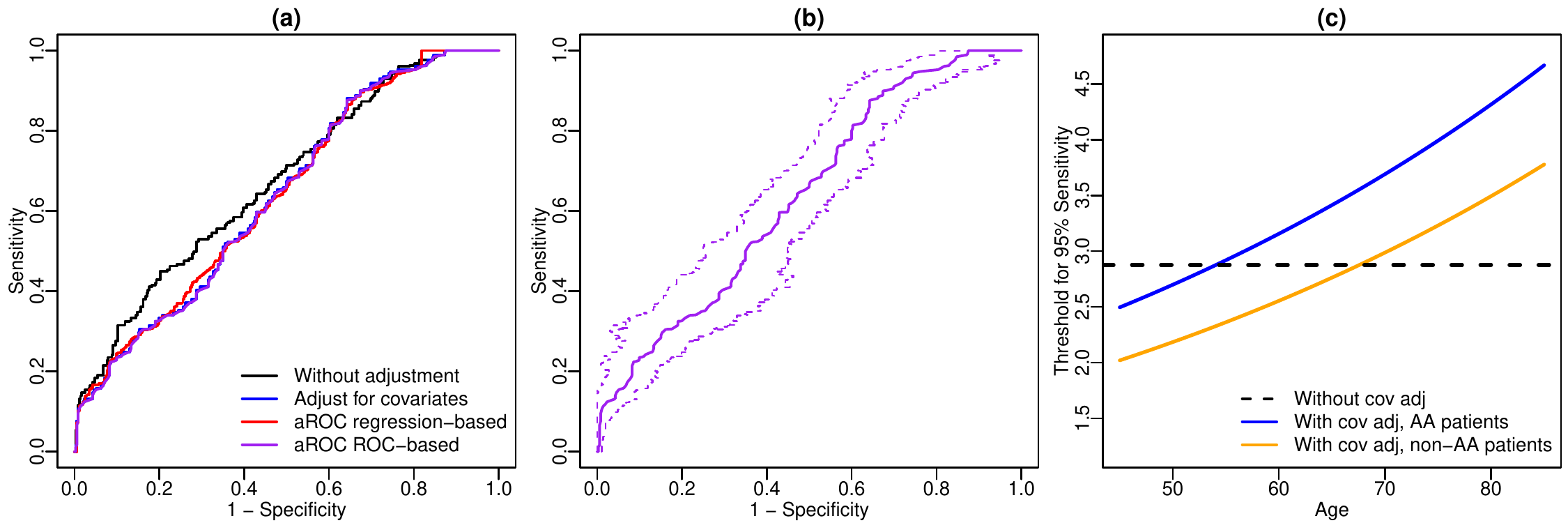}
	\caption{\textbf{ROC curve and threshold results for the prostate study.} Panel (a): ROC curve without and with adjustment for covariates (black and blue curve, respectively). Red and purple curves are adjusted ROC curves after applying regression- and ROC-based monotonization. Panel (b): Covariate-adjusted ROC curve with ROC-based monotonization is in solid purple. Dashed purple lines are the $95\%$ confidence band.  Panel (c): Estimated PSA threshold at controlled $95\%$ sensitivity level by age based on the local model.}
	\label{real2}
\end{figure}

\end{document}